\documentclass[aps,prc,reprint,amsmath,amssymb,nofootinbib,preprintnumbers]{revtex4-1}

\usepackage[pdfpagelabels=false]{hyperref}
\hypersetup{
	colorlinks,
	linkcolor={blue},
	citecolor={blue},
	urlcolor={blue}
}

\def\MK{{\mathcal{M}_K}}    
\def\MQ{{\mathcal{M}_Q}}   
\def\MKS{{\mathcal{S}_K}}  
\def\KQ{{K_Q}}                       

\def\dd{\mathrm{d}}
\def\SU{\operatorname{SU}}
\def\U{\operatorname{U}}
\def\Re{\operatorname{Re}}
\def\Im{\operatorname{Im}}

\def\norm#1{\left|\!\left| {#1} \right|\!\right|}

\begin{document}
\preprint{MITP--24--042}
\title{On the Geometry of $\mathcal{N}=2$ Minkowski Vacua of \\[0.5ex]
Gauged $\mathcal{N}=2$ Supergravity Theories in Four Dimensions}
\author{Hans Jockers}
\email{jockers@uni-mainz.de}
\author{S\"oren Kotlewski}
\email{soeren-kotlewski@t-online.de}
\affiliation{
PRISMA+ Cluster of Excellence \& Mainz Institute for Theoretical Physics\\
Johannes Gutenberg-Universit\"at Mainz\\
55099 Mainz, Germany
}
\begin{abstract}
Gauging isometries of four-dimensional $\mathcal{N}=2$ supergravity theories yields an $\mathcal{N}=2$ supersymmetric theory with a scalar potential. In this note, we study the well-known constraints for four-dimensional $\mathcal{N}=2$ Minkowski vacua of such theories. We propose that classically a projective special K\"ahler submanifold of the projective K\"ahler target space of the ungauged theory describes the moduli space of the complex scalar fields of massless vector multiplets for $\mathcal{N}=2$ Minkowski vacua configurations, which then receives quantum corrections from integrating out massive fields. Subloci of projective special K\"ahler manifolds appear as supersymmetric flux vacua in the context of type~IIB Calabi--Yau threefold compactifications with background fluxes as well. While these flux vacua equations arise from the critical locus of an $\mathcal{N}=1$ superpotential, we show that these equations can also be obtained from the $\mathcal{N}=2$ supersymmetric Minkowski vacuum equations of gauged $\mathcal{N}=2$ supergravity theories upon gauging suitable isometries in the semi-classical universal hypermultiplet sector of type~IIB string Calabi--Yau threefold compactifications. Thus, we give an intrinsic $\mathcal{N}=2$ supersymmetric interpretation to the flux vacua equations. 
\end{abstract}
\maketitle


\section{Introduction} \label{sec:intro}
It is well-established that the low-energy effective action of type~IIB string theory compactified on a smooth Calabi--Yau threefold is an ungauged four-dimensional $\mathcal{N}=2$ supergravity theory with a moduli space of $\mathcal{N}=2$ supersymmetric Minkowski vacua \cite{Bohm:1999uk,Bodner:1989cg}. Such ungauged $\mathcal{N}=2$ supergravity theories consist of a single gravity multiplet, $n_v$ vector multiplets, and $n_h$ hypermultiplets \cite{deWit:1984rvr,Andrianopoli:1996cm}. The scalar fields of the vector multiplets and the hypermultiplets parametrize a projective K\"ahler manifold $\MK$ of complex dimension $n_v$ and a quaternionic K\"ahler manifold $\MQ$ of real dimension $4 n_h$, respectively.  In the context of Calabi--Yau compactifications the projective K\"ahler target space manifold $\MK$ is identified with the complex structure moduli space of the Calabi--Yau threefold \cite{Cecotti:1989kn,Ceresole:1992su,Bershadsky:1993cx}, whereas a semi-classical approximation of the quaternion K\"ahler manifold $\MQ$ is obtained from the universal hypermultiplet and the complexified quantum K\"ahler moduli space of complex dimension $(n_h-1)$ of the Calabi--Yau threefold via the c-map \cite{Cecotti:1988qn,Ferrara:1989ik,Andrianopoli:1996bq}. 

In this work, we want to study for Calabi--Yau threefolds  with higher-dimensional complex structure moduli spaces $\MK$ that possess lower-dimensional projective K\"ahler submoduli spaces $\MKS \subset \MK$. Such projective K\"ahler submoduli spaces $\MKS$ furnish again suitable target spaces of ungauged four-dimensional $\mathcal{N}=2$ supergravity theories with $\dim_\mathbb{C} \MKS$ vector multiplets. Instead of considering a $\mathcal{N}=2$ supergravity theory with the vector multiplet target space $\MKS$ directly, we want to realize the target space $\MKS$ as the space of $\mathcal{N}=2$ supersymmetric Minkowski vacua of a scalar potential in the vector multiplet sector arising from a gauged four-dimensional $\mathcal{N}=2$ supergravity theory \cite{deWit:1984rvr,Andrianopoli:1996cm}.

Examples of projective special K\"ahler submanifolds~$\MKS$ are extremal transition loci in the complex structure moduli space of Calabi--Yau threefolds \cite{Greene:1995hu,Katz:1996ht,Klemm:1996kv}. Other examples appear for families of Calabi--Yau threefolds with enhanced discrete symmetries, such as the complex structure moduli space $\MKS$ of the one-parameter Dwork family of quintic Calabi--Yau threefolds with the $(\mathbb{Z}_5)^3$ Greene--Plesser symmetry \cite{Greene:1990ud}, which is embedded in the complex structure moduli space $\MK$ of the $101$-parameter family of generic quintic Calabi--Yau threefolds.

The motivation for studying projective special K\"ahler submanifolds from the $\mathcal{N}=2$ supergravity perspective is two-fold. Firstly, we suggest a physical mechanism that allows us to localize in the infrared to  a projective K\"ahler target submanifold $\MKS \subset \MK$ of the large vector multiplet target space manifold $\MK$. Secondly, such a construction is motivated by string compactifications on Calabi--Yau threefolds with background fluxes. Flux vacua of string compactifications on Calabi--Yau threefolds --- as for instance recently studied with modern arithmetic methods in refs.~\cite{Kachru:2020abh,Kachru:2020sio,Schimmrigk:2020dfl,Candelas:2023yrg} --- are expected either to be truncated $\mathcal{N}=1$ Calabi--Yau orientifold compactifications \cite{Brunner:2003zm,Grimm:2004uq} or to correspond to $\mathcal{N}=2$ gauged supergravity theories \cite{Mayr:2000hh,Louis:2002ny,Grana:2006hr,DAuria:2007axr,Samtleben:2008pe}. It is the latter scenario that we want to entertain here from the supergravity perspective.

The conditions for supersymmetric flux vacua in Calabi--Yau threefolds are typically formulated as the critical locus of a flux induced superpotential, which per se is a notion of $\mathcal{N}=1$ supersymmetric theories \cite{Taylor:1999ii,Mayr:2000hh,Grimm:2004uq}. These flux vacua are also closely related to the supersymmetric attractors in the context of supersymmetric black hole solutions \cite{Moore:1998zu,Moore:1998pn}, which can be given a Hodge-theoretic formulation in the context of complex structure moduli spaces of Calabi--Yau threefolds. The main result of this work is that the flux vacua equations --- which arise from the critical locus of an $\mathcal{N}=1$ superpotential --- can also be obtained from gauging a pair of isometries in the universal hypermultiplet sector of $\mathcal{N}=2$ supergravity theories of type~IIB Calabi--Yau threefold compactifications, which is a manifest $\mathcal{N}=2$ supersymmetric construction.  

\section{Subloci of Projective Special K\"ahler Manifolds} \label{sec:Sublocus}
To describe the projective special K\"ahler target space manifold $\MK$ of an ungauged $\mathcal{N}=2$ supergravity theory with $n_v$ vector multiplets, we consider the $2(n_v +1)$-dimensional (complex) vector space $V$ with the real symplectic basis $(\alpha_I,\beta^J)$, $I, J= 0,\ldots,n_v$, and the symplectic skew-symmetric pairing $\langle \alpha_I , \beta^J \rangle = \delta^{J}_I$, $\langle \alpha_I , \alpha_J \rangle = 0$, $\langle \beta^I , \beta^J \rangle = 0$. 
The vector space $V$ together with its the canonical $\mathbb{C}^*$-action preserving the symplectic structure admits a $\mathbb{C}^*$-equivariant holomorphic Lagrangian immersion of a conical special K\"ahler manifold $\mathcal{M}_{\text{cK}} \subset V$ of complex co-dimension one. Then the quotient space $\MK = \mathcal{M}_{\text{cK}} /{\mathbb{C}^*}$ gives rise to a projective special K\"ahler manifold $\MK$ of complex dimension $n_v$. The details of this constructions are developed in refs.~\cite{MR1894078,Cortes:2011aj}.\footnote{See also ref.~\cite{Freed:1997dp} for an equivalent construction of a projective special K\"ahler manifold.}

Locally, the projective special K\"ahler manifold~$\MK$ is characterized by the complex vector
\begin{equation}
  \Omega(X^0,\ldots, X^{n_v}) = X^I \alpha_I - F_J \beta^J \ .
\end{equation}  
Here $X^I$, $I = 0,\ldots,n_v$, are the complex projective coordinates and $F_J$, $J=0,\ldots,n_v$, are the derivatives of the holomorphic prepotential~$F(X^0,\ldots,X^{n_v})$ given by
\begin{equation}
  F_I = \frac{\partial F}{\partial X^I} \ .
\end{equation}
The holomorphic prepotential~$F(X^0,\ldots,X^{n_v})$ is of homogeneous degree two, i.e., 
\begin{equation}
  F(\lambda X^0,\ldots,\lambda X^{n_v}) = \lambda^2 \, F(X^0,\ldots, X^{n_v}) \ .
\end{equation} 
Geometrically, the zero locus of the degree two prepotential~$F(X^0,\ldots,X^{n_v})$ describes locally the $\mathbb{C}^*$-equivariant Lagrangian immersion of the conical special K\"ahler manifold $\mathcal{M}_{\text{cK}} \hookrightarrow V$ \cite{Cortes:2011aj}.

The imaginary part of the complex symmetric matrix $F_{IJ} = \frac{\partial^2 F}{\partial X^I \partial X^J}$ has signature $(n_v,1)$ \cite{Ceresole:1995ca}. Furthermore, the real and positive K\"ahler potential $K$ of the projective special K\"ahler manifold $\MK$ reads
\begin{equation}
  K = - \log i \langle \overline\Omega, \Omega \rangle \ ,
\end{equation}
which gives rise to a positive definite Hermitian K\"ahler metric
\begin{equation}
  G_{i\bar\jmath}(z,\bar z) = \frac{\partial^2 K}{\partial z^i \partial\bar z^{\bar\jmath}} \ , \qquad i,j=1,\ldots,n_v \ ,
\end{equation}
in terms of a choice of affine local coordinates
\begin{equation}
  z^i = \frac{X^i}{X^0} \ , \qquad i=1,\ldots, n_v \ .
\end{equation}    

We consider now a projective special K\"ahler submanifold $\MKS$ of complex dimension $s$, $1\le s < n_v$. Let $S\subset V$ be a $2(s+1)$-dimensional symplectic subvector space of the symplectic vector space $V$, which we assume without loss of generality to be spanned by the basis vectors $(\alpha_A,\beta^B)$, $A,B=0,\ldots,s$. Furthermore, the affine coordinates $z^i = (x^a, y^m)$, $a=1,\ldots,s$, $m=1,\ldots, n_v -s$, are given by
\begin{equation}
  x^a = \frac{X^a}{X^0} \ , \qquad y^m = \frac{X^{s+m}}{X^0} \ .
\end{equation}  
Then we define a projective special K\"ahler submanifold $\MKS$ as the sublocus $y^m=0$, if for all $x^a$
\begin{equation} \label{eq:FCon1}
    \left.F_{AJ}\right|_{y^m=0} = 0 \ , \quad A=0,\ldots,s \ , \   J=s+1,\ldots, n_v \ ,
\end{equation}
and the imaginary parts of the $(s+1)\times(s+1)$-matrix $\left.F_{ab}\right|_{y^m=0}$, $a,b=0,\ldots,s$, and the $(n_v-s)\times(n_v-s)$-matrix  $\left.F_{s+m,s+n}\right|_{y^m=0}$, $m,n=1,\ldots,n_v-s$, are non-degenerate with signatures $(s,1)$ and $(n_v-s,0)$, respectively. Then the complex vector $\Omega$ restricted to $y^m=0$ yields the K\"ahler potential $K^\MKS$ 
\begin{equation}
  K^\MKS(x^1,\ldots, x^s) =  \left. -\log i \langle \overline\Omega, \Omega \rangle \right|_{y^m=0} \ ,
\end{equation}  
which is the local K\"ahler potential of the projective special K\"ahler submanifold $\MKS$. Note that the conditions~\eqref{eq:FCon1} are equivalent to
\begin{equation} \label{eq:FCon2}
  X^{s+m} = 0 \ , \qquad   F_{s+m}(X^0,\ldots,X^s,0,\ldots,0) = 0 \ ,
\end{equation}  
for all $m=1,\ldots,n_v-s$. This means that the symplectic pairs consisting of the homogeneous coordinates $X^{s+m}$ and derivatives of the prepotential $F_{s+m}$ have to vanish along the submanifold $\MKS$. 

In the context of compactifications of type~IIB string theory on a Calabi--Yau threefold~ the projective special K\"ahler manifold $\MK$ is the complex structure moduli space of the Calabi--Yau threefold. The symplectic basis $(\alpha_I,\beta^J)$ is identified with a basis of three-form cohomology classes generating $H^3(X,\mathbb{R})$ and its symplectic structure arises from the skew-symmetric intersection pairing on $H^3(X,\mathbb{R})$. The complex vector $\Omega$ becomes the nowhere vanishing holomorphic $(3,0)$-form generating the one-dimensional Dolbeault cohomology class $H^{(3,0)}(X)$, and --- as a consequence of Griffiths' transversality \cite{MR229641,MR233825,MR282990} --- the derivatives $\partial_{I_1}\cdots\partial_{I_k}\Omega(X^0,\ldots,X^{n_v})$ of $\Omega(X^0,\ldots,X^{n_v})$ with respect to the projective coordinates $X^I$, $I=0,\ldots,n_v$, for arbitrary but finitely many $I_1,\ldots,I_k$ span the complex cohomology group $H^3(X,\mathbb{C}) = H^3(X,\mathbb{R}) \otimes \mathbb{C}$, i.e.,
\begin{equation}
  H^3(X,\mathbb{C}) =  \langle\!\langle \, \partial_{I_1} \cdots\partial_{I_k}\Omega(X^0,\ldots,X^{n_v}) \,
    \rangle\!\rangle \ .
\end{equation}
A projective special K\"ahler subspace $\MKS$ of the entire complex structure moduli space $\MK$ is now characterized in terms of Griffiths' transversality by the property that the space of cohomology classes $S$ is given by
\begin{equation}
  S = \left. \langle\!\langle \,  \partial_{A_1} \cdots\partial_{A_k}\Omega(X^0,\ldots,X^{n_v})\, \rangle\!\rangle\right|_{y^m=0}   \subset H^3(X,\mathbb{C}) \ .
\end{equation}
Here the derivatives $\partial_{A_i}$ are taken with respect to the first $(s+1)$ projective coordinates $X^{A_i}$,  $0,\ldots,s$, only. If $\dim_\mathbb{C}S < \dim_\mathbb{C} H^3(X,\mathbb{C})$, $S$ furnishes a sub-Hodge structure of $H^3(X,\mathbb{C})$. Note that this is a highly non-generic condition, because taking success derivatives of $\Omega$ along an arbitary coordinate direction $z^i$ at any random point in the complex structure moduli space generically generates the entire three-form cohomology. Namely, generically we have $H^3(X,\mathbb{C}) \simeq \langle\!\langle \Omega,\partial_{z^i}\Omega, \partial_{z^i}^2\Omega, \partial_{z^i}^3\Omega,\partial_{z^i}^4\Omega, \ldots \rangle\!\rangle$.

Note that the arithmetic attractor loci introduced in refs.~\cite{Moore:1998pn,Moore:1998zu} realize non-trivial zero-dimensional subloci in the vector multiplet target space, which in our setup can be viewed as zero-dimensional projective special K\"ahler submanifolds. These attractor loci enjoy a physical interpretation in the context of the attractor mechanism, which describes black hole solutions of $\mathcal{N}=2$ supergravity \cite{Ferrara:1995ih}.

\section{4d $\mathcal{N}=2$ Supergravity Theories} \label{sec:SUGRA}
Our aim is now to construct a four-dimensional $\mathcal{N}=2$ supergravity theory with the projective special K\"ahler manifold $\MK$ as its target space in the vector multiplet sector together with a scalar potential $V$ that dynamically constrains the flat directions of the scalar fields in the vector multiplet sector to the projective special K\"ahler submanifold $\MKS\subset\MK$. In other words, we want to construct an $\mathcal{N}=2$ supergravity theory with target space $\MK$, whose moduli space of supersymmetric $\mathcal{N}=2$ Minkowski vacua is parametrized in the vector multiplet sector by the submanifold $\MKS$. 

As opposed to four-dimensional ungauged $\mathcal{N}=1$ supergravity theories, which admit a holomorphic superpotential of the $\mathcal{N}=1$ chiral multiplets resulting in a scalar potential, the four-dimensional ungauged $\mathcal{N}=2$ supergravity theories cannot have a scalar potential for any of their scalar fields. As a consequence, all scalar fields of ungauged $\mathcal{N}=2$ supergravity theories parametrize flat directions of a real $(2n_v + 4n_h)$-dimensional moduli space of four-dimensional $\mathcal{N}=2$ Minkowski vacua. This moduli space factors into $\MK \times \MQ$ \cite{deWit:1984rvr,Andrianopoli:1996cm}, where $\MK$ is the projective special K\"ahler moduli space of complex dimension $n_v$ of the vector multiplet sector, and $\MQ$ is the quaternionic K\"ahler moduli space of the hypermultiplet sector. As a consequence, it is not possible to lift the flat directions of the scalar degrees of freedom within the framework of ungauged $\mathcal{N}=2$ supergravity theories. In particular, it is impossible to constrain with effective ungauged $\mathcal{N}=2$ supergravity theories the projective special K\"ahler target space $\MK$ of the vector multiplet scalar fields to a submanifold $\MKS$.

However, a scalar potential is generated in gauged $\mathcal{N}=2$ supergravity theories \cite{deWit:1984rvr,Andrianopoli:1996cm,Mayr:2000hh,Hristov:2009uj,Louis:2012ux}. Let $z^i$, $i=1,\ldots,n_v$, be the complex scalar fields of the vector multiplets, $q^u$, $u=1,\ldots,4 n_H$, the real scalar fields of the hypermultiplets, and $A_\mu^I$, $I=0,\ldots,n_v$, the $(n_v+1)$-electric gauge fields of the graviphoton in the gravity multiplet and of the gauge fields in the $n_v$ vector multiplet. 

In order to describe gaugings of magnetic gauge fields as well, we follow refs.~\cite{Hristov:2009uj,Louis:2012ux} and consider in addition to the electric vector fields their dual magnetic gauge fields $B_{\mu,J}$ \cite{Louis:2012ux}, as proposed in ref.~\cite{Mayr:2000hh}. For ease of notation, we combine the electric and the magnetic vector fields into the $2(n_v+1)$ vector fields $(C^\Lambda_\mu) = (A_\mu^I,B_{\mu,J})$, $\Lambda=1,\ldots,2(n_v+1)$. Similarly, we pair the projective special K\"ahler coordinates $X^I$ with their derivatives of the prepotential $F_J$ into $(Z^\Lambda) = (X^I, F_J)$, $\Lambda=1,\ldots,2(n_v+1)$. We call the quantities $Z^\Lambda$ --- which are all of homoegenous degree one with respect to the projective coordinates $X^I$ --- the periods of the projective special K\"ahler target space $\MK$. 

Let us now assume that the target spaces $\MK$ and $\MQ$ of the scalar fields in the vector and hypermultiplet sectors possess continuous symmetries, which in turn give rise to Killing vector fields $k^i_\lambda(z) \partial_i$ in $\MK$ and $\tilde k^u_{\tilde\lambda}(q) \partial_u$ in $\MQ$. Here the indices $\lambda$ and $\tilde\lambda$ label the symmetries of $\MK$ and $\MQ$, respectively. We arrive at an $\mathcal{N}=2$ gauged supergravity theory upon gauging (some of) these isometries by introducing the gauge covariant derivatives for the vector multiplet scalar fields \cite{deWit:1984rvr,Andrianopoli:1996cm,Hristov:2009uj,Louis:2012ux}
\begin{equation}
  D_\mu z^i = \partial_\mu z^i -C_\mu^\Lambda k^i_\Lambda(z) \ , \quad
  k_\Lambda^i(z) =  \Theta^\lambda_\Lambda k^i_\lambda(z) \ ,
\end{equation}
and for the hypermultiplet scalar fields  
\begin{equation}
  D_\mu q^u = \partial_\mu q^u -C_\mu^\Lambda \tilde{k}^u_{\Lambda}(q) \ , \quad
  \tilde{k}^u_{\Lambda}(q)(q) = \tilde{\Theta}^{\tilde\lambda}_\Lambda \tilde{k}^u_{\tilde{\lambda}}(q) \ .
\end{equation} 
In these gauge covariant derivatives the constants $\Theta_\Lambda^\lambda$ and $\tilde{\Theta}_\Lambda^{\tilde{\lambda}}$ denote the embedding tensor to the gauge fields \cite{Louis:2012ux}, which represents the choice of representation for the gauge group of the vector- and hypermultiplets, respectively. The vector fields $k^i_\Lambda(z) \partial_i$ and  $\tilde{k}^u_{\Lambda}(q) \partial_u$ with index $\Lambda=1,\ldots,2(n_v+1)$, denote the Killing vectors that appear in the covariant derivatives as governed by their embedding tensors.

Since the scalar fields $z^i$ reside in the vector multiplet, they must transform in the adjoint representation with respect to the gauge symmetry \cite{Hristov:2009uj,Louis:2012ux}. This imposes strong constraints on the vector multiplet embedding tensor $\Theta^\lambda_\Lambda$, which also implies that the scalar fields $z^i$ can never be gauged for Abelian vector multiplets.

In order for the resulting theory to be $\mathcal{N}=2$ supersymmetric, additional terms appear in the gauged $\mathcal{N}=2$ supergravity Lagrangian. In particular, the gauged $\mathcal{N}=2$ supergravity theory possesses the scalar potential \cite{deWit:1984rvr,Andrianopoli:1996cm,Mayr:2000hh,Hristov:2009uj,Louis:2012ux}
\begin{multline} \label{eq:Vpot}
  V(z,q) = e^{K(z)}\! \left[2
    \norm{\overline{Z}^\Lambda(z) k^i_\Lambda(z) }^2_\MK\!\!\!\!\!\!
    + 4\norm{\overline{Z}^\Lambda(z) \tilde k^u_{\Lambda}(q)}^2_\MQ\right. \\
     \left. + \operatorname{tr}  \norm{\nabla^i \overline{Z}^\Lambda(z) \mathcal{P}_{\Lambda}(q)}^2_\MK\!\!\!\!\!\!
    -\frac{3}{2}\operatorname{tr}  \left|  Z^\Lambda(z)  \mathcal{P}_{\Lambda}(q) \right|^2
    \right] \ . 
\end{multline}   
Here $\norm{\,\cdot\,}_\MK$ and $\norm{\,\cdot\,}_\MQ$ are the norms of the special K\"ahler metric $g_{i\bar\jmath}(z) = \partial_i\partial_{\bar\jmath} K(z)$ of $\MK$ and the quaternionic K\"ahler metric $g_{uv}(q)$ of $\MQ$, $\nabla_i = \partial_i  + K_i(z)$ is the K\"ahler covariant derivative of $\MK$, and $\mathcal{P}_\Lambda(q) = ( \mathcal{P}^a_\Lambda(q))$, $a=1,2,3$, is the triplet of $\mathfrak{su}(2)$ Lie algebra valued Killing prepotentials and the trace $\operatorname{tr}(\,\cdot\,)$ refers to the positive definite bilinear Killing form of the $\mathfrak{su}(2)$ Lie algebra acting on the Lie-algebra valued Killing prepotentials $\mathcal{P}_\Lambda$. The Killing prepotentials $\mathcal{P}_\Lambda(q)$ obey
\begin{equation} \label{eq:KPrepots}
   -2\tilde{k}^u_\Lambda(q) K_{uv}^a(q) = \nabla_v\mathcal{P}_\Lambda^a(q) \ , \quad a=1,2,3 \ .
\end{equation}
Here $\nabla_v \mathcal{P}_\Lambda^a=\partial_v\mathcal{P}_\Lambda^a+\epsilon^{abc}\omega^{b}_v\mathcal{P}_\Lambda^c$ is the $\operatorname{SU}(2)$-covariant derivative with respect to the subgroup $\operatorname{SU}(2) \simeq \operatorname{Sp}(1)$ of the holonomy $\operatorname{Sp}(n)\cdot\operatorname{Sp}(1)$ of the quanternionic K\"ahler manifold~$\MQ$, and $K^a = d\omega^a + \frac12 \epsilon^{abc} \omega^b \wedge \omega^c$ is the curvature of the connection $\omega^{a}_v$. For more details on the scalar potential $V(z,q)$, see for instance refs.~\cite{deWit:1984rvr,Andrianopoli:1996cm,Mayr:2000hh,Hristov:2009uj,Louis:2012ux}.

As the cosmological constant vanishes in a Minkowski vacuum, it is necessary that the scalar potential vanishes as well. Except for the last term in the scalar potential~\eqref{eq:Vpot}, all remaining contributions  are non-negative. It is further shown in refs.~\cite{Hristov:2009uj,Louis:2012ux} that for an $\mathcal{N}=2$ supersymmetric Minkowski vacuum, the non-positive term in the scalar potential $V(z,q)$ must also vanish by itself. Therefore, altogether we have that for an $\mathcal{N}=2$ supersymmetric Minkowski vacua the conditions
\begin{equation} \label{eq:SUSYConst}
\begin{aligned}
	0&=Z^\Lambda(z) \, \mathcal{P}_{\Lambda}(q) \ , \quad &
	0&=\nabla_{\bar{\jmath}}\bar{Z}^\Lambda(z) \, \mathcal{P}_{\Lambda}(q) \ , \\
	0&=\overline{Z}^\Lambda(z) \, \tilde{k}_{\Lambda}^u(q) \ , \quad &
	0&=\overline{Z}^\Lambda(z) \, k^i_{\Lambda}(z)  \  
\end{aligned}
\end{equation}
need to be obeyed. The constraint $0=\overline{Z}^\Lambda(z) \, k^i_{\Lambda}(z)$ gives a relation only among the vector multiplet scalars, whereas the constraints from the gauging of the isometries on the quaternionic K\"ahler manifold realize interactions between the vector multiplets and the hypermultiplets. Upon inserting the differential equation~\eqref{eq:KPrepots} for the Killing vectors $\tilde{k}_{\Lambda}^u(q)$, these three relations can be equivalently formulated as
\begin{equation} \label{eq:SUSYConst2}
\begin{aligned}
	0&=Z^\Lambda(z) \, \mathcal{P}_{\Lambda}(q) \ , \quad &
	0&=Z^\Lambda(z) \, \partial_v \mathcal{P}_{\Lambda}(q) \ , \\
	0&=\partial_{i} Z^\Lambda(z) \, \mathcal{P}_{\Lambda}(q) \ .
\end{aligned}
\end{equation}
These conditions are indeed equivalent since the triplet of curvature two-form $K^a_{uv}$ are invertible for any $a=1,2,3$. This can be seen by noting that $K^a_{uv}(q)$ can be expressed in terms of the quaternionic K\"ahler metric $g_{uw}(q)$ and the triplet of almost complex structures $J^a(q)$, $a=1,2,3$, on $\MQ$ as $K^a_{uv}(q)=g_{uw}(q)(J^a(q))^w_v$ \cite{Andrianopoli:1996cm}.

\section{Space of $\mathcal{N}=2$ Minkowski Vacua} \label{sec:Vacua}

Given a solution to the $\mathcal{N}=2$ Minkowski vacuum equations~\eqref{eq:SUSYConst} in terms of expectation values of the scalar fields $z$ and $q$, the deformations to these expectation values preserving eqs.~\eqref{eq:SUSYConst} correspond to flat directions of the scalar potential~\eqref{eq:Vpot} and give rise to $\mathcal{N}=2$ massless multiplets. The remaining obstructed deformations of the scalar fields --- not in accordance with the $\mathcal{N}=2$ Minkowski vacuum equations~\eqref{eq:SUSYConst} --- generically assemble themselves into massive $\mathcal{N}=2$ multiplets \cite{Louis:2012ux}. The obtained low energy effective theory of the massless $\mathcal{N}=2$ multiplets furnishes an $\mathcal{N}=2$ supergravity theory of massless fields with a projective special K\"ahler and a quanternionic K\"ahler target space for the vector multiplets and the hypermultiplets, respectively. We propose that semi-classically the vector multiplet target space is a projective special K\"ahler submanifold~$\MKS$ of the target space $\MK$ of the ungauged supergravity theory. At the quantum level the geometry~$\MKS$ receives one-loop perturbative and non-perturbative quantum corrections from integrating out the massive $\mathcal{N}=2$ multiplets \cite{Seiberg:1994rs,Seiberg:1994aj}.

More specifically, let us now discuss the possible different gaugings and their resulting $\mathcal{N}=2$ Minkowski vacuum structure. We distinguish between gaugings of isometries in the projective special K\"ahler target space manifold $\MK$ and of isometries in the quanternionic K\"ahler target space manifold $\MQ$. 

As discussed in refs.~\cite{Hristov:2009uj,Louis:2012ux}, gauging isometries of the projective special K\"ahler manifold are constrained such that the scalar field $z^i$ can only transform non-trivially in the adjoint representation of a non-Abelian compact gauge group. For generic loci, where the induced non-negative term $\norm{\overline{Z}^\Lambda(z) k^i_\Lambda(z) }^2_\MK$ in the scalar potential~\eqref{eq:Vpot} vanishes, the non-Abelian gauge group is broken to its maximal torus and the moduli space of the $\mathcal{N}=2$ supersymmetric Minkowski vacuum realizes the Coulomb branch of the non-Abelian $\mathcal{N}=2$ gauge theory coupled to gravity. For non-generic $\mathcal{N}=2$ supersymmetric Minkowski vacua there can still be an unbroken non-Abelian gauge subgroup. For the various strata in the Coulomb branch the Higgs mechanism generates a mass for the broken gauge fields, which combine with the massive scalar fields into short massive BPS vector multiplets \cite{Louis:2012ux}. Assuming that integrating out these massive BPS vector multiplets yields again an effective $\mathcal{N}=2$ supersymmetric supergravity theory in terms of a Lagrangian description, the massless $\mathcal{N}=2$ vector multiplets are again governed by a projective special K\"ahler target space manifold of smaller dimension than $\MK$. The obtained effective prepotential~$F$ is not simply a classical reduction of the prepotential of the original target space $\MK$, but in addition it receives a one-loop correction and further non-perturbative instanton corrections from integrating out the massive multiplets~\cite{Seiberg:1994rs}.

The remaining terms in the scalar potential~\eqref{eq:Vpot} of gauged $\mathcal{N}=2$ supergravity theories stem from gaugings of isometries of the quanternionic K\"ahler manifold $\MQ$. For $\mathcal{N}=2$ supersymmetric Minkowski vacua these gaugings impose the remaining three types of constraints~\eqref{eq:SUSYConst2}, which involve both the scalar fields from the vector and the hypermultiplet sector of the $\mathcal{N}=2$ gauged supergravity theory. As a consequence, constraining the projective special K\"ahler manifold $\MK$ to a submanifold $\MKS$ in this way requires a quaternionic special K\"ahler manifold with suitable isometries. We discuss these gaugings and their resulting $\mathcal{N}=2$ Minkowski vacua in the context of $\mathcal{N}=2$ supergravity theories arising form Calabi--Yau threefold compactifications in the next section.

\section{Gauged Type~IIB Calabi--Yau Threefold Compactifications} \label{sec:3fold}
The low energy effective action of type~IIB string theory compactified on a Calabi--Yau threefold yields an ungauged four-dimensional $\mathcal{N}=2$ supergravity theory  \cite{Bohm:1999uk,Bodner:1989cg}. In such compactifications, the complex structure moduli of the Calabi--Yau threefold realize the $n_V$ vector multiplets, and the Calabi--Yau K\"ahler moduli give rise to $n_H-1$ hypermultiplets that combine with the universal hypermultiplet (containing the dilaton) to the $n_H$ hypermultiplets of the $\mathcal{N}=2$ supergravity theory. 

To discuss gaugings of such a hypermultiplet sector, we need to have a handle on the quaternionic K\"ahler manifold $\MQ$ for such compactifications. The structure of the quaternionic K\"ahler manifold from Calabi--Yau compactifications is of a very special type and can be constructed semi-classically via the c-map from the complex structure moduli space of the mirror Calabi--Yau manifold \cite{Cecotti:1988qn,Ferrara:1989ik,Andrianopoli:1996bq}. For any Calabi--Yau threefold compactification of type~IIB string theory, the resulting quaternionic K\"ahler manifolds always contain the universal hypermultipet sector, whose scalar fields correspond to the complex axio-dilaton and (the duals of) the complex two-dimensional two-from tensor field arising from the $B$-field and the Ramond--Ramond two-form field. The remaining $n_H-1$ hypermultiplets are comprised of the complexified K\"ahler moduli of the Calabi--Yau threefold  and the internal $B$-field and the Ramond--Ramond two-form fields of the compactification Calabi--Yau threefold. The semiclassical quaternionic target spaces $\MQ$ contstructed via the c-map exhibit a rich structure of isometries \cite{Ferrara:1989ik}, which can be gauged. On the quantum level, the semi-classical quaternionic target space geometry receives intricate corrections that are challenging to compute, see for instance the reviews~\cite{Alexandrov:2013yva,Alexandrov:2011va}.

In this note, we focus on gaugings of the semiclassical universal hypermultiplet, and leave the discussion for other gaugings to future work.\footnote{More general gaugings of the hypermultiplet sector of the low energy effective action of M-theory compactified on a Calabi--Yau threefold  are studied in refs.~\cite{Brandle:2002fa,Jarv:2003qx,Lukas:2004du,Mohaupt:2004pq,Mohaupt:2005pa}. These five-dimensional low energy effective supergravity theories relate to the  four-dimensional $\mathcal{N}=2$ gauged supergravity theories discussed in this note via further dimensional reduction on a circle.}
Our motivation for considering gaugings in the universal hypermultiplet sector is two-fold: On the one hand, the universal hypermultiplet does not depend on the geometry of the specific Calabi--Yau threefold.  Hence, the gaugings of the universal hypermultiplet sector are applicable to any Calabi--Yau threefold compactification of type~IIB string theory. On the other hand and more importantly for us, gaugings of the universal hypermultiplet sector are closely related to the flux vacua equations of Calabi--Yau geometries recently analyzed in refs.~\cite{Kachru:2020abh,Kachru:2020sio,Schimmrigk:2020dfl,Candelas:2023yrg} by modern arithmetic techniques. Thus, the goal of the remainder of this section is to exhibit a connection between such flux vacua and the universal hypermultiplet gaugings of four-dimensional $\mathcal{N}=2$ gauged supergravity theories.

The $\mathcal{N}=2$ flux vacuum equations arise from the critical locus of the flux generated $\mathcal{N}=1$ superpotential $W$ of the form \cite{Gukov:1999ya,Taylor:1999ii}
\begin{equation}\label{eq:Wflux}
  W(z,\tau) = \int (F - \tau H) \wedge \Omega(z) \ ,
\end{equation}  
where $H$ and $F$ are the Neveu--Schwarz and Ramond--Ramond background three-form fluxes, $\tau$ is the complex axio-dilaton, and $\Omega$ is the holomorphic $(3,0)$ form of the Calabi--Yau threefold. This superpotential is given by a semi-classical analysis and receives perturbative and non-perturbative quantum corrections \cite{Witten:1996bn,Gukov:1999ya,Kachru:2003aw,Berglund:2005dm,Jockers:2009ti}. Spelled out in refs.~\cite{Mayr:2000hh,Kachru:2020abh,Kachru:2020sio,Schimmrigk:2020dfl,Candelas:2023yrg} the critical locus of flux vacua yield the constraints
\begin{equation} \label{eq:Wconst}
\begin{aligned}
   &Z^\Lambda(z) f_\Lambda = 0 \ , \qquad
   Z^\Lambda(z) h_\Lambda = 0 \ ,  \\
   &(\partial_i Z^\Lambda(z)) (f_\Lambda - \tau h_\Lambda) = 0 \ .
\end{aligned}   
\end{equation}
Here the (in suitable units) rational coefficients $f_\Lambda$ and $h_\Lambda$ are the flux quanta of the  three-form background fluxes $F$ and $H$, respectively. The first two equations arise from the requirement that the superpotential and the derivative with respect to the axio-dilaton vanish in a flux vacuum, whereas the last equation is obtained from the requirement that the gradient of the superpotential $W$ with respect to the complex structure moduli $z^i$ ought to vanish as well. In the context of $\mathcal{N}=2$ supergravity theories, the flux-induced superpotential~\eqref{eq:Wflux} relates to a complex linear combination of two real components of the triplet of the Killing prepotentials $\mathcal{P}_\Lambda$ \cite{Berglund:2005dm,Mayr:2000hh}. Due to the prominent appearance of the axio-dilaton in the flux vacuum equations~\eqref{eq:Wconst}, it is natural to consider the Killing prepotentials attributed to the universal hypermultiplet.

The quaternionic geometry of the universal hypermultiplet is well-studied, see for instance refs.~\cite{SFerrara_1989,Ketov:2001gq,Strominger:1997eb,Cortes:2020klb}, and it can be described in terms of the coset space $\SU(2,1)/\operatorname{S}(\U(2)\times \U(1))$. As opposed to a generic quaternionic K\"ahler manifold, which is not K\"ahler, the semi-classical universal quaternionic K\"ahler manifold is actually a K\"ahler manifold, whose complex local coordinates are the complex coordinate $C$ associated to the two-form tensors of the universal hypermultiplet and the complex coordinate $S$, which reads
\begin{equation}
   S = e^{-\phi} + i \sigma + C \overline{C} \ .
\end{equation}
Here $\phi$ and $\sigma$ are the real dilaton and the real axion of the universal hypermultiplet. In terms of the complex coordiantes the K\"ahler potential $\KQ$ of the universal hypermultiplet then takes the form
\begin{equation}
  \KQ = - \log \left[ S + \overline{S} - 2 C \overline{C} \right] \ ,
\end{equation}  
which results in the K\"ahler metric
\begin{multline}
  \dd s^2 = e^\KQ \left( \dd S \dd\overline{S} - 2 C \dd S \dd \overline{C} - 2 \overline{C} \dd\overline{S} \dd C \right. \\
  \left. + 2 (S +\overline{S}) \dd C\dd\overline{C} \right) \ . 
\end{multline}
The continuous isometries of the quaternionic K\"ahler manifold of the universal hypermultiplet constitute of the real shift symmetry $\sigma \to \sigma + s$, $s\in \mathbb{R}$, the $\U(1)$ rotation of the complex variable $C$, and the symmetry $C \to C + \epsilon$, $S \to S + 2 C \bar\epsilon + \epsilon^2$, $\epsilon \in \mathbb{C}$. Altogether, these four real isometries yield the respective four real Killing vectors
\begin{equation} \label{eq:kUHM}
\begin{aligned}
  \tilde k_{1} &= i\left(\partial_S - \partial_{\overline{S}} \right) \ ,  \qquad
  \tilde k_{2} = \tfrac{i}{2}\left(C\partial_C - \overline{C} \partial_{\overline{C}} \right) \ , \\
  \tilde k_{3} &=\tfrac{1}{2}(\partial_C + \partial_{\overline{C}}) -  \tfrac{i}{2} \Im C  \left(\partial_S - \partial_{\overline{S}} \right) \ , \\
  \tilde k_{4} &= -\tfrac{i}{2} \left( \partial_C - \partial_{\overline{C}} \right) + \tfrac{i}{2} \Re C  \left(\partial_S - \partial_{\overline{S}} \right) \ .
\end{aligned}  
\end{equation}
Solving the Killing vector equation~\eqref{eq:KPrepots}, we arrive at the associated real $\mathfrak{su}(2)$-valued Killing prepotentials
\begin{align}
   \mathcal{P}_{1} &= \tfrac{1}2 e^{\phi} i\sigma^3 \ , \nonumber \\
   \mathcal{P}_{2} &= -e^{\phi/2} \left( \Re C  i\sigma^1 +  \Im{C} i\sigma^2 \right)  
      +\tfrac12 (1 - e^{\phi} C \overline{C}) i\sigma^3  \ , \nonumber \\ 
   \mathcal{P}_{3} &= e^{\phi/2} i\sigma^2 +e^{\phi} \Im{C} i\sigma^3 \ ,  \nonumber \\
   \mathcal{P}_{4} &= e^{\phi/2} i\sigma^1 +e^{\phi} \Re{C} i\sigma^3 \ ,  \label{eq:PUHM}
\end{align}   
where the generators of the Lie algebra $\mathfrak{su}(2)$ are given by $i\sigma^a$, $a=1,2,3$, with the Pauli matrices $\sigma^a$. 

Let us now construct a gauged $\mathcal{N}=2$ supergravity theory that makes contact with the flux vacua equations~\eqref{eq:Wconst}. We pick two independent isometries of the universal hypermultiplet sector that correspond to two Killing prepotentials $\mathcal{P}_{(1)}(S,C)$ and $\mathcal{P}_{(2)}(S,C)$, which are functions of the complex fields $S$ and $C$. With respect to these two isometries we gauge the vector multiplets by choosing the embedding tensor $\tilde\Theta^{\tilde\lambda}_\Lambda$ such that the Killing prepotentials contracted with the embedding tensor become
\begin{equation} \label{eq:Ptwoiso}
  \mathcal{P}_\Lambda(S,C) = f_\Lambda \mathcal{P}_{(1)}(S,C) - h_\Lambda \mathcal{P}_{(2)}(S,C) \ , 
\end{equation}  
in terms of the flux quanta $f_\Lambda$ and $h_\Lambda$. Recall that the Killing prepotentials $\mathcal{P}_{(1)}(S,C)$ and $\mathcal{P}_{(2)}(S,C)$ take values in the Lie algebra $\mathfrak{su}(2)$. As a result for independent isometries and for generic expectation values of the scalar fields $S$ and $C$, these two Killing prepotentials realize linearly independent $\mathfrak{su}(2)$ Lie algebra elements. As a consequence, the $\mathcal{N}=2$ Minkowski vacuum constraints~\eqref{eq:SUSYConst2} yield, for generic expectation values of $S$ and $C$, the equations
\begin{equation} \label{eq:GenericN2}
\begin{aligned}
    Z^\Lambda(z)f_\Lambda &= 0 \ , \quad &Z^\Lambda(z)h_\Lambda &= 0 \ , \\
    \partial_i Z^\Lambda(z)f_\Lambda &= 0 \ , \quad &\partial_iZ^\Lambda(z)h_\Lambda &= 0 \ .
\end{aligned}    
\end{equation}
These constraints are more restrictive than the flux vacuum equation~\eqref{eq:Wconst} because --- due to the linearly-independent prepotentials~$\mathcal{P}_{(1)}(S,C)$ and $\mathcal{P}_{(2)}(S,C)$ ---  the gradients of the periods $Z^\Lambda(z)f_\Lambda$ and $Z^\Lambda(z)h_\Lambda$ are forced to vanish separately.  However, for the flux vacua equations~\eqref{eq:Wconst} only the linear combination of the gradients --- as governed by the expectation value of the axio-dilaton --- must be zero.

However, the comparison of the generic vacua conditions~\eqref{eq:GenericN2} with the flux vacua equations~\eqref{eq:Wconst} does not involve the same number of degrees of freedom because the $\mathcal{N}=2$ universal hypermultiplet depends on the expectation value of four real scalar fields, whereas the $\mathcal{N}=1$ complex axio-dilaton $\tau$ consists only of two real scalar fields. Therefore, we impose the condition that the expectation values of the scalar fields $S$ and $C$ are restricted such that Lie algebra valued prepotentials $\mathcal{P}_{(1)}(S,C)$ and $\mathcal{P}_{(2)}(S,C)$ become linearly dependent.  That is to say the Lie algebra valued Killing prepotentials $\mathcal{P}_{(1)}(S,C)$ and $\mathcal{P}_{(2)}(S,C)$ viewed as three-dimensional real vectors become parallel in the Lie algebra~$\mathfrak{su}(2)$, i.e., 
\begin{equation}
  (S,C) \in \mathcal{T} \ , \qquad
  \mathcal{T} = \left\{ (S,C) \, \middle|\, \mathcal{P}_{(1)} \parallel \mathcal{P}_{(2)} \right\} \ .
\end{equation}  
We expect that the space $\mathcal{T}$ of such expectation values is of real dimension two because the alignment of two three-dimensional vectors requires two real degrees of freedom out of the four real degrees of freedom of the universal hypermultiplet. Thus this condition matches the two real degrees of freedom of the axio-dilaton $\tau$ in the flux vacua equations~\eqref{eq:Wconst}. Therefore, we call this condition the axio-dilaton non-genericity constraint. It implies that eq.~\eqref{eq:Ptwoiso} restricted to the expectation values of $\mathcal{T}$ becomes
\begin{equation} \label{eq:ADcond}
  \left.\mathcal{P}_\Lambda(S,C)\right|_{\mathcal{T}} = \left(f_\Lambda - \tau_\mathcal{T} \, h_\Lambda\right)
  \left. \mathcal{P}_{(1)}(S,C) \right|_{\mathcal{T}} \ ,
 \end{equation}
where $\tau_\mathcal{T}$ is a function of the non-generic expectation values $(S,C)$ in the set $\mathcal{T}$ of the axio-dilaton non-genericity constraint. Note, however, the gradient of eq.~\eqref{eq:Ptwoiso} with respect to the universal hypermultiplet fields $S$ and $C$ restricted to $\mathcal{T}$ still remains a sum of two linearly independent Lie algebra valued quantities because the Killing prepotentials $\mathcal{P}_{(1)}$ and $\mathcal{P}_{(2)}$ are by assumption associated to two independent isometries. As a result, by imposing the axio-dilaton non-genericity constraint, we arrive at the $\mathcal{N}=2$ Minkowski vacua equations 
\begin{equation}
\begin{aligned}
    &Z^\Lambda(z)f_\Lambda = 0 \ , \quad Z^\Lambda(z)h_\Lambda = 0 \ , \\
    &\partial_i Z^\Lambda(z)\left(f_\Lambda -\tau_\mathcal{T} h_\Lambda\right) = 0  &\text{for} \ (S,C)\in\mathcal{T} \ .
\end{aligned}    
\end{equation}
These equations agree with the flux vacua equations~\eqref{eq:Wconst} upon identifying the $\mathcal{N}=1$ axio-dilaton field $\tau$ with the constrained hypermultiplet function $\tau_\mathcal{T}$ for the non-generic expectation values of $(S,C)\in \mathcal{T}$. 

Let us illustrate this class of gauging with an explicit choice of universal hypermultiplet isometries that is given in terms of the Killing prepotentials of eq.~\eqref{eq:PUHM} by 
\begin{equation}
\mathcal{P}_{(1)}=\mathcal{P}_{1} \ , \ \mathcal{P}_{(2)}=\mathcal{P}_{2} \ .
\end{equation} 
Hence, we can read off directly that $\mathcal{P}_{(1)}\parallel\mathcal{P}_{(2)}$ if and only if $\Re(C)=\Im(C)=0$ and we we conclude that 
\begin{equation}
	\mathcal{T}=\{(S,0) \ | \ S\in\mathbb{C} \} \ .
\end{equation}
As proposed in the general discussion, this implies that only two real degrees of freedom of the universal hypermultiplet remain unconstrained along the axio-dilaton non-genericity locus $\mathcal{T}$. Moreover, on this space of non-generic expectation values, we find 
\begin{equation}
	\left.\mathcal{P}_2\right|_{\mathcal{T}}=e^{-\phi}\left.\mathcal{P}_1\right|_{\mathcal{T}} \ ,
\end{equation}
such that eq.~\eqref{eq:ADcond} reduces to
\begin{equation}
 \left.\mathcal{P}_\Lambda(S,C)\right|_{\mathcal{T}} = \left(f_\Lambda + e^{-\phi} \, h_\Lambda\right)
  \left. \mathcal{P}_{(1)}(S,C) \right|_{\mathcal{T}} \ .
\end{equation}
Thus, for this choice of gauging and the non-generic choice of expectation values $(S,C)\in\mathcal{T}$, the $\mathcal{N}=2$ Minkowski vacuum constraints eq. (\ref{eq:SUSYConst2}) are realized by
\begin{equation}
\begin{aligned}
    &Z^\Lambda(z)f_\Lambda = 0 \ , \quad Z^\Lambda(z)h_\Lambda = 0 \ , \\
    &\partial_i Z^\Lambda(z)\left(f_\Lambda - e^{-\phi} h_\Lambda\right) = 0 \ .
\end{aligned}    
\end{equation}
In contrast to the previous discussion we obtain that the constrained hypermultiplet function~$\tau_\mathcal{T}=e^{-\phi}$ does not encode both remaining degrees of freedom of the universal hypermultiplet, but only the real dilation $\phi$. The axion~$\sigma$ is unconstrained by these vacuum conditions. 

The property that the constrained hypermultiplet function~$\tau_\mathcal{T}$ is independent of the real axion $\sigma$ is not specific to the considered choices of Killing prepotentials $\mathcal{P}_{(1)}$ and $\mathcal{P}_{(2)}$ in this explicit example. Instead, it is a consequence of the shift symmetry of $\sigma$, which at the classical level prohibits a functional dependence of $\tau_\mathcal{T}$ on the real axion $\sigma$ --- also for any other two choices of a pair of isometries. However, upon truncating to the $\mathcal{N}=1$ setting, the field dependent function $\tau_\mathcal{T}$ must always become a holomorphic function of two real scalar degrees of freedom because the $\mathcal{N}=1$ superpotential is a holomorphic function of $\mathcal{N}=1$ chiral fields. We expect that the non-generic dependence of the function $\tau_\mathcal{T}$ on a single real degree of freedom does not occur once quantum corrections are taken into account. 

\section{Discussion and Conclusions} \label{sec:con}
In this note, we consider the interplay between gauged isometries of the target spaces of $\mathcal{N}=2$ gauged supergravities and the resulting semi-classical spaces of $\mathcal{N}=2$ Minkowski vacua, which are the critical loci of the gauged $\mathcal{N}=2$ supergravity theories. We propose that --- after integrating out all massive degrees of freedom in such $\mathcal{N}=2$ Minkowski vacua and under the assumption that the remaining massless degrees of freedom enjoy a Lagrangian description --- the scalar fields of the massless $\mathcal{N}=2$ vector multiplets parametrize a projective special K\"ahler target space, which arises as a quantum deformation of a submanifold of the projective special K\"ahler manifold that is assoicated to the ungauged $\mathcal{N}=2$ supergravity theory.

We focus on those four-dimensional $\mathcal{N}=2$ supergravity theories which are obtained as low energy effective theories of type~IIB string compactifications. Such supergravity theories possess a universal hypermultiplet sector, and we study explicitly gauging isometries of this sector. We show that the critical locus of the $\mathcal{N}=1$ flux-induced superpotential of Calabi--Yau threefolds arises also from the $\mathcal{N}=2$ vacuum equations of the supergravity theory by gauging two independent isometries in the universal hypermultiplet sector. 

Our motivation for studying the gauged $\mathcal{N}=2$ supergravity theories obtained from the isometries of the universal hypermultiplet sector is that such gaugings do not depend on the specific details of the chosen Calabi--Yau threefold compactification space. Nevertheless, we believe that gauging more general quaternionic isometries is an interesting research direction to pursue. In particular,  we expect that extremal transitions between topologically distinct Calabi--Yau threefolds in the context of type~IIB string compactifications are realized in terms of gauged $\mathcal{N}=2$ supergravity theories, in which both projective special K\"ahler and quaternionic K\"ahler isometries beyond the universal hypermultiplet sector are gauged. For such gauged supergravity theories the space of $\mathcal{N}=2$ Minkowski vacua realizes in the Higgs branch the projective special K\"ahler submanifold, which is a submanifold of the projective special K\"ahler manifold of the Coulomb branch. In the field theory limit, the interesting works~\cite{Greene:1995hu,Katz:1996ht,Klemm:1996kv} discuss in detail the connection between the geometric Calabi--Yau extremal transitions and their realization in terms of type~IIB string theory compactifications. Formulating these extremal transitions in the context of an effective gauged $\mathcal{N}=2$ supergravity description promises an interesting interplay between the projective special K\"ahler and the quanternionic K\"ahler manifolds of the vector- and hypermultiplet sectors beyond the field theory limit discussed in refs.~\cite{Greene:1995hu,Katz:1996ht,Klemm:1996kv}.\footnote{For transitions between Calabi--Yau threefolds in the context of M-theory compactifications from the five-dimensional supergravity perspective, see refs.~\cite{Brandle:2002fa,Jarv:2003qx,Mohaupt:2004pq}.}

Finally, let us remark that stringy quantum corrections to the low-energy effective $\mathcal{N}=2$ theories of Calabi--Yau compactifications are expected to break the target space isometries of the ungauged $\mathcal{N}=2$ supergravity theories. We believe that there is an interplay between such quantum corrections and the gauging of isometries along the lines of ref.~\cite{Kashani-Poor:2005eiu}, where the gauging of isometries modifies the zero mode structure of symmetry breaking instantons. A detailed understanding of the relationship between non-perturbatively broken target space isometries and gauged $\mathcal{N}=2$ supergravity theories possibly reveals a non-trivial interplay between quantum effects in the vector multiplet and in the hypermultiplet sector, which may, for instance, have geometric implications in enumerative geometry for pairs of Calabi--Yau threefolds that are connected via extremal transitions.

\section*{Acknowledgements}
We thank Andreas Morr for collaborating at early stages of this project.
We are grateful to Pyry Kuusela for many illuminating discussions and for his detailed comments on a first version of this manuscript. 
We would like to thank
Aravind Aikot,
Sergey Ketov
and
Peter Mayr
for interesting discussions and useful correspondences.
This work is supported by the Cluster of Excellence Precision Physics, Fundamental Interactions, and Structure of Matter (PRISMA+, EXC 2118/1) within the German Excellence Strategy (Project-ID 390831469).

\bibliographystyle{apsrev4-1} 
\bibliography{JK.bib}
\end{document}